# Standards in the Preparation of Biomedical Research Metadata: A Bridge2AI Perspective

J. Harry Caufield[1], Satrajit Ghosh[2,3], Sek Wong Kong[4], Jillian Parker[6], Nathan Sheffield[7,8,9,10], Bhavesh Patel[10], Andrew Williams[†,11,12], Timothy Clark[7], and Monica C. Munoz-Torres[13]

[1] Division of Environmental Genomics and Systems Biology, Lawrence Berkeley National Laboratory, Berkeley, CA 94720, USA
[2] McGovern Institute for Brain Research, Massachusetts Institute of Technology, Cambridge, MA 02139, USA
[3] Department of Otolaryngology, Harvard Medical School, Boston, MA 02115, USA
[4] Computational Health Informatics Program, Boston Children's Hospital, Boston, MA 02215, USA
[5] Department of Pediatrics, Harvard Medical School, Boston, MA 02115, USA
[6] University of California San Diego School of Medicine, La Jolla, CA 92093, USA
[7] Department of Public Health Sciences, School of Medicine, University of Virginia, Charlottesville, VA 22908, USA
[8] Department of Biomedical Engineering, School of Medicine, University of Virginia, Charlottesville, VA 22904, USA
[9] Department of Biochemistry and Molecular Genetics, School of Medicine, University of Virginia, Charlottesville, VA 22908, USA
[10] California Medical Innovations Institute, San Diego, CA 92121, USA
[11] Clinical and Translational Science Institute, Tufts Medical Center, Boston, MA 02111, USA
[12] Institute for Clinical Research and Health Policy Studies, Tufts Medical Center, Boston, MA 02111, USA
[13] Department of Biomedical Informatics, University of Colorado Anschutz Medical Campus, Aurora, CO 80045, USA
† In memoriam

## Background

AI-readiness describes the degree to which data may be optimally and ethically used for subsequent AI and Machine Learning (AI/ML) methods, where those methods may involve some combination of model training, data classification, and ethical, explainable prediction. The Bridge2AI consortium has defined the particular criteria a biomedical dataset may possess to render it AI-ready: in brief, a dataset's readiness is related to its FAIRness, provenance, degree of characterization, explainability, sustainability, and computability, in addition to its accompaniment with documentation about ethical data practices.[1] Biomedical datasets present specific challenges for AI-readiness. They may only rarely be reliably treated as "ground truths", are often extensively pre-processed, may be derived from human subjects with legal and other limitations on their use, require special treatment to provide pre-model explainability, and have other characteristics that make them subject to unique criteria.

Metadata is "data that describes or gives information about other data."[2] Data without context or provenance is meaningless and has been cited as an especially vexing problem in many AI applications, affecting findability, determination of data authenticity, consent, provenance, reproducibility, ethics, and other issues.[3,4] Metadata needs to be machine- and human-readable, and mapped to standardized vocabularies or ontologies to ensure its interoperability.[5,6] These criteria are established via standards, which may apply globally, across a research domain, or for a particular project or program. Within Bridge2AI, standardization regarding the internet, Web, and biomedical research information standards enables the outputs to be conveniently found, widely and reliably used, and to meet AI-readiness criteria.

To ensure AI-readiness and to clarify data structure and relationships within Bridge2AI's Grand Challenges (GCs), particular types of metadata are necessary. The GCs within the Bridge2AI initiative include four data-generating projects focusing on generating AI/ML-ready datasets to tackle complex biomedical and behavioral research problems. These projects develop standardized, multimodal data, tools, and training resources to support AI integration, while addressing ethical data practices. Examples include using voice as a biomarker, building interpretable genomic tools, modeling disease trajectories with diverse multimodal data, and mapping cellular and molecular health indicators across the human body. For more details about the consortium's work, please visit [https://bridge2ai.org](https://bridge2ai.org).

## Motivation and Scope

Managing metadata presents challenges for many biomedical research projects. Researchers may lack understanding of applicable standards or the necessary tools. Bespoke standards may exist without mechanisms for interoperability, leading to standardization becoming a downstream task accomplished by curators or secondary data integration projects. In cases where clearly defined standards exist, they may lack a clear mechanism for making metadata compatible with other data sets.

We formed a *Standards for Metadata And Project Structure* (SMAPS) sub-working group within Bridge2AI, intending to identify current practices in the GCs regarding how project-level metadata is created, stored, and communicated. We used the efforts of Bridge2AI data generators as examples, and our observations are representative of Bridge2AI research. Accordingly, our observations and discussions span the wide range of data modalities and sources within this consortium, from multi-omics studies focused on individual cellular components to multi-modal clinical studies from multiple sites. Our focus was specifically on creating standardized project-level metadata, which may include descriptions of experimental methods, details of results, consent statements, and/or contexts for observed clinical phenotypes. This scope includes metadata for individual data tables and records describing larger data sets.

The motivation for this report is to assess the state of metadata creation and standardization in the Bridge2AI GCs, provide guidelines where required, and identify gaps and areas for improvement across the program. A major objective of all the Bridge2AI GC projects is that all data released is pre-standardized to the extent required for appropriate use by domain-knowledgeable AI researchers. New projects, including those outside the Bridge2AI consortium, would benefit from what we have learned about creating metadata as part of efforts to promote AI readiness.

Our working group reviewed the studies performed within all four GCs in meetings held between July 2024 and January 2025. These discussions covered the experiences and perspectives of representatives from each GC. Rather than prescribing certain standards or processes for researchers to prepare metadata, we inventoried the tools and platforms in active use within the consortium. We then identified features of these resources that contribute to AI-readiness.

Project-level metadata plays a crucial role because a single approach may not capture all necessary metadata for every use case. Some metadata properties can apply to individual data components, while others are specific to particular data types. For instance, in a multi-omics study, samples might originate from the same source but undergo different parallel analyses, like transcriptomics and proteomics. Ideally, metadata standards should represent both the origin of the data (provenance) and the metadata relevant to each specific data type, possibly through combining existing standards.

## Project Metadata Standards in Bridge2AI Grand Challenges

Here, we discuss the standards employed by each of the four Bridge2AI Grand Challenges: AI/ML for Clinical Care, Functional Genomics, Precision Public Health, and Salutogenesis. Each is a multi-site program collecting multiple types of biomolecular and/or biomedical data, all with the overarching goal of providing their data sets in an AI-ready form. An overview of this section is provided in **Figure 1**.

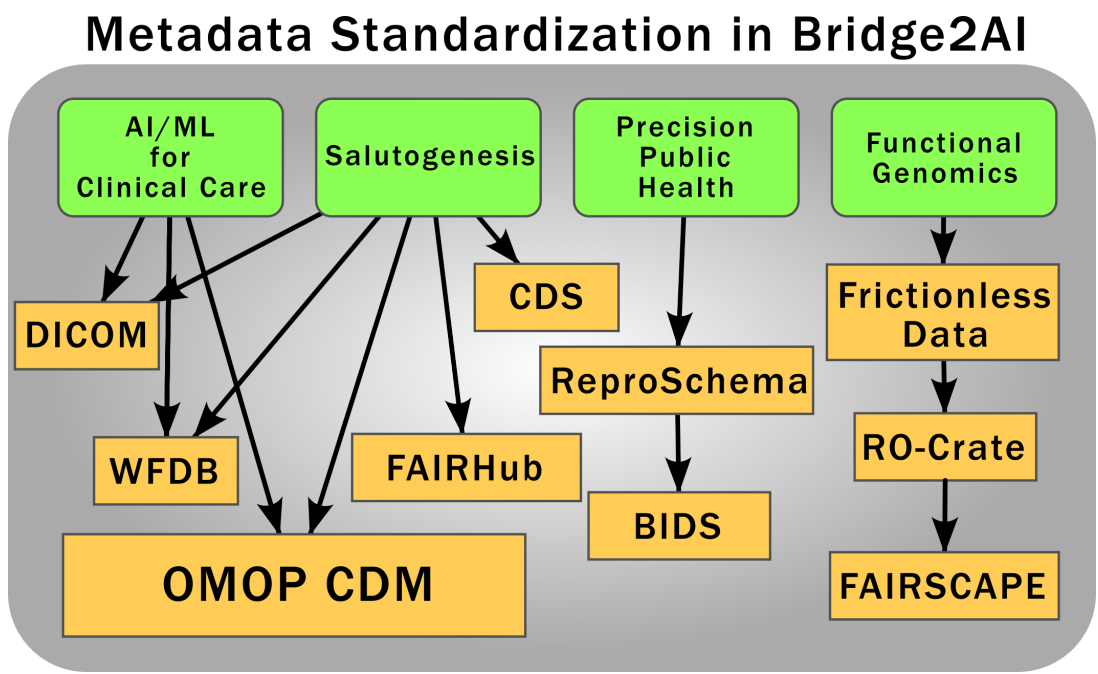

**Fig. 1. Overview of specific metadata tools and standards employed by Bridge2AI Grand Challenge projects.** BIDS, Brain Imaging Data Structure; CDS, Clinical Dataset Structure; DICOM, Digital Imaging and Communications in Medicine; OMOP CDM, Observational Medical Outcomes Partnership (OMOP) Common Data Model; WFDB, WaveForm DataBase format. This figure is not comprehensive; please see text for details.

### *AI/ML for Clinical Care*

This GC, also known as Patient-Focused Collaborative Hospital Repository Uniting Standards (CHoRUS) for Equitable AI, uses and extends standards for intensive care clinical data to support analyses illuminating recovery from critical illness. It collects a wide variety of data, including medical waveforms (e.g., from electrocardiograms), images, notes, questionnaires, health-related geospatial data, and various types of tabular electronic health record (EHR) data. Data sets collected by the AI/ML for Clinical Care GC currently include millions to billions of rows of standardized data from some individual sites in different modalities and varying contributions from 15 medical centers. New records are added to its central enclave as they become available. The dataset is currently accessible in a secure enclave. Permissioned users interrogate and analyze the data using a mature suite of tools that cover the full range of observational study workflows.

The overall model used to unify these data types is an extension of version 5.4 of the OMOP common data model (CDM).[7] File formats are standardized for waveforms (Waveform Database Software Package)[8] and imaging (Digital Imaging and Communications in Medicine [DICOM®, ISO 12052:2017][9]). CHoRUS work motivated specific additions to the OMOP CDM. This includes the specification and addition of a new Vocabulary Metadata table for tracking details about the terms and names used in a given dataset. This table will be introduced in the CDM's next versioned release. The GC is also developing official extensions to the Standard OMOP Vocabulary to accommodate new data types. These are intended to ensure consistency with OMOP and extend the scope of OHDSI tooling by continuous engagement with OHDSI standards and tool developers. CHoRUS also built an extensible public curation process for semantic mapping for valuable but costly to acquire data elements and tooling for partially automating the extraction, transformation, and loading (ETL) of these data into a standardized form.[10] These new standards are being validated in public beta testing before the broader version 1.0 release of the dataset and publication of the new or extended standards.

The schema for OMOP fully specifies all the data in each version of a dataset conforming to a versioned release. CHoRUS uses version 5.4. The conventions for mapping data to the schema are specified by the THEMIS project.[11] CHoRUS has also investigated how to best apply the FAIRScape platform to transparently document data provenance. They have specifically used metadata and vocabulary for place-based data as a focused use case. Further details are available in the OHDSI Geographic Information System (GIS) repository.[12]

Standardized metadata are used differently for different purposes in OHDSI and CHoRUS. Persistent unique identifiers for millions of medical concepts are used to annotate instances of healthcare records in a relational structure designed for longitudinal analyses. Separate databases within the OMOP CDM contain standardized results in a schema and vocabulary for summarized aggregate person-level information. Other schemas contain the vocabularies used to annotate the data. Still others contain person-level results for derived time-span units. Finally, all of CDM's components, including its tables, fields, vocabulary concepts, and relationships, are represented in the CDM as unique standardized concepts. This use of metadata has advantages in methods and software development.

Data quality in the CHoRUS GC is assessed through standardized scripts that check conformance, completeness, and plausibility at table, variable, and concept levels. This results in thousands of detailed query results organized by data quality and rendered as Structured Query Language (SQL) code. Procedures for local and central data quality are documented in standard operating procedures (SOPs). Integrated datasets are released as beta versions before full standards validation. ID management involves a complex strategy for local alignment across data modalities and central integration using a registry to ensure reproducible patient IDs. CHoRUS-developed extensions to vocabulary and schema are used to inform OMOP standards development and are in draft form until validated and integrated into official releases, as are expert-validated semantic mappings.

### *Functional Genomics*

This GC, also known as Cell Maps for AI (CM4AI), has the objective of mapping the spatiotemporal architecture of human cells and using the resulting maps to learn more about relationships between cellular genotypes and phenotypes. It produces machine-readable maps of cell architecture as AI-ready data resulting from multimodal processing of 100 chromatin modifiers and 100 metabolic enzymes, all from cell lines relevant to important disease states. Cells are observed under perturbed and unperturbed conditions. This GC is also developing reusable toolsets and frameworks for producing cell maps.

CM4AI uses the Research Object Crate (RO-Crate)[13] with JSON-LD serialization to package its datasets, with data described using Schema.org, DataCite, Evidence Graph Ontology (EVI)[14,15], JSON Schema, and Frictionless Data metadata vocabularies. Within packages, data formats used include FASTQ and HDF5 for perturb-seq data, Thermo Fisher's RAW for mass spectroscopy data, and both JPEG and OME-TIFF[16] for subcellular imaging data. Network data for cell maps are described using the CX format, which can be uploaded to the Network Data Exchange (NDEx) for storage, sharing, manipulation, and publication.[17,18]

CM4AI assigns persistent identifiers (PID) to all datasets and software it releases, which are packaged using the RO-Crate standard, as described above. CM4AI's FAIRSCAPE clients create RO-Crates with JSON-LD serialized Schema.org metadata[19] extended with terms from additional relevant ontologies as needed, including references to dataset schemas for each dataset and provenance information. Cell line sample IDs are specified in the dataset descriptions. All provenance graphs are resolvable in the FAIRSCAPE server via ARK[20] PIDs to component datasets and archived software versions. Wherever possible, software archived in GitHub is also archived in Zenodo to receive a DOI and associated DataCite schema metadata[21]. Provenance graphs are provided to connect data files with specific experiment and sample identifiers; these follow the EVI model. CM4AI addresses sample characterization in metadata by integrating the Portable Encapsulated Projects (PEP) approach[22] with its provenance graphs.

The GC provides data dictionaries as Frictionless Data[23] schemas on each dataset in the current CM4AI RO-Crate packages. These schemas define the dataset structure, column labels, and descriptions where relevant, datatypes, and mappings to appropriate vocabulary terms if they exist. Validation code is also provided. In some cases, such as image files, the data are not columnar. Current practice in CM4AI is to provide schemas specifying the MIME type and any constraints, such as dimensionality and colorset, for validation. More complex cases, such as OME-TIFF and DICOM images, are expected to be encountered where part of the dataset consists of defined metadata and the rest of the pixels or voxels. In such cases, CM4AI plans to implement data dictionaries for the required metadata. RO-Crate packages contain a reference to, and are validated against, the latest RO-Crate version specification.

### *Precision Public Health*

This GC, also known as Voice as a Biomarker of Health, is specifically concerned with using recordings of human voice to find potential connections to disease states. They seek to identify health biomarkers to promote voice AI research and build computational models to assist in screening, diagnosis, and treatment. Their work has produced a set of more than 16,000 recordings from 442 participants, as well as open-source code for processing and working with voice data at different levels of anonymization. Participants are separated into distinct disease cohorts: in addition to a control group, there are cohorts for voice disorders (including a pediatric cohort), respiratory diseases, mood/psychiatric disorders, and neurological/neurodegenerative disorders. Data is collected through a smartphone application designed specifically for this project.

The Precision Public Health GC uses an organization schema aligned with the Brain Imaging Data Structure (BIDS)[24] as a packaging model for its datasets. Audio data is stored in WAV format, while clinical and phenotypic form data is stored as tab-separated values (TSV) with JSON data dictionaries. All files have a corresponding JSON file containing key metadata. Derivatives are generated to enable AI/ML ready interaction. The waveform features are stored using two formats: 1) A fixed feature format that includes static features extracted from the entire waveform, and 2) A temporal format that varies for each audio file depending on the length of recording. To ensure broader data interoperability, the team has also developed a Voice as a Biomarker for AI Health profile for FHIR R4, derived from the US Core STU5 profile.

Research performed by the Precision Public Health GC also produces a wealth of data from patient surveys. This includes answers to questions about demographics and habits with potential health impact (e.g., smoking and drinking habits) as well as an extensive array of questionnaires specific to each disease cohort. For example, participants in the voice disorder cohort are asked to complete the Voice Handicap Index-10 (VHI-10), Patient Health Questionnaire (PHQ-9), and General Anxiety Disorder (GAD-7) surveys.

All raw audio and questionnaire data is stored in a REDCap system, then converted to the BIDS structure. Data dictionaries are generated using software tools that organize the data and extract features from the audio. These dictionaries retrieve information from the structured protocols in ReproSchema format[25] and from the SenseLab audio feature extraction toolkit; SenseLab is an open-source Python package developed in the course of the GC's work in Bridge2AI.[26] The dictionaries are provided in BIDS JSON format.

*Salutogenesis*

This GC, also known as Artificial Intelligence Ready and Equitable Atlas for Diabetes Insights (AI-READi), is focused on creating an ethically-sourced dataset of Type 2 Diabetes (T2D) patients. It has yielded a large and multimodal set of clinical observations, with more than one thousand participants and 13 types of measurements spanning from survey data and physical metrics to retinal images and environmental properties (e.g., humidity and temperature at participants' homes).

As of its v2.0.0 release, the AI-READI dataset is organized and packaged according to the Clinical Dataset Structure (CDS), a set of standard practices they have engineered to structure a dataset consistently.[27] The CDS is inspired by the same Brain Imaging Data Structure (BIDS) adapted by the Precision Public Health GC. Individual data modalities use the following formats: Waveform Database (WFDB) format for cardiac echocardiogram data; OMOP CDM for clinical observations; Earth Science Data System (ESDS) for environmental data (specifically, the ASCII File Format Guidelines for Earth Science Data[28]); DICOM for retinal imaging; and the Open mHealth standards[29] for wearable sensor data. For clinical data, data dictionaries are provided in the dataset documentation.

The Salutogenesis GC assigns a unique DOI to each dataset version released through the FAIRhub platform. The DOI and all provenance metadata are included with the dataset following the CDS. The metadata, packaged across different metadata files, links to all relevant IDs, including the ClinicalTrials.gov ID of the study, funding ID, the ORCID of the Principal Investigators, the ROR ID of their affiliations and of the study sites, and ID of relevant keywords from controlled vocabularies and ontologies. The dataset documentation provides a detailed description of each data type, including how it was collected, processed, and formatted to support reproducibility.

## Storing Metadata in Repositories

The National Institutes of Health (NIH) defines a set of recommended repositories in their guidelines for sharing data. Most are specific to particular research data domains. Some others are more general, such as the repositories included in NIH's Generalist Repository Ecosystem Initiative (GREI). These are recommended for depositing data with no obvious home in another, more domain-specific archive.

Bridge2AI GCs producing data containing personal health information have implemented strategies for appropriate protection and governance. The AI/ML for Clinical Care GC uses a common date shifting method and pixel scrubbing to limit re-identification risks. Though the GC excludes patient location, it provides regional health attributes. Similarly, the Precision Public Health GC employs methods to anonymize its voice data, including through novel approaches developed specifically for the project; initial surveys found that few consistent standards existed for collecting voice data, regardless of strategies for retaining patient privacy.[30] The Precision Public Health GC made an initial data release without raw audio waveforms or personally identifying information on the Health Data Nexus repository[31], a platform maintained by the

Temerty Centre for Artificial Intelligence Research and Education in Medicine (T-CAIREM) at the University of Toronto. Access is via terms of a restrictive, PHI-protecting license, to properly identified researchers only. A separate release was made through Physionet[32], maintained by the teams at the Massachusetts Institute of Technology Laboratory for Computational Physiology and the Margret and H.A. Rey Institute for Nonlinear Dynamics at Beth Israel Deaconess Medical Center. Future releases will implement governance principles for access to audio waveforms. Salutogenesis GC data releases are made through the FAIRhub platform as described above. FAIRhub is designed to provide access to datasets while minimizing privacy risk. Prospective data users must agree to a custom license and use an identity verification system.

The Functional Genomics GC releases its data and metadata on NIH-approved domain-specific and generalist repositories, targeting domain-specific repositories as the default where they exist for a particular domain. Where domain-specific repositories do not exist, or where data acquisition labs have not been able to arrange deposition for various reasons, the GC deposits data in generalist repositories that participate in the NIH GREI. An example of this is shown by how the GC handles its pertub-seq data[33]: Illumina base calls are deposited in the NIH's Sequence Read Archive (SRA), a domain-specific repository, in FASTQ format, while data emerging from an analysis pipeline including steps for pre-processing, quality control, batch correction, is structured as HDF5 (h5) files. These HDF5 files are deposited in the general-purpose Figshare repository, which the perturb-seq community has agreed to use for this type of experiment, given the lack of a domain-specific repository dedicated to this purpose. Outputs of the Functional Genomics GC's integration pipeline are deposited in the University of Virginia's implementation of the open source Dataverse platform[34], with provenance graphs referencing the identifiers of pipeline inputs in their designated repository. Cell map outputs are additionally deposited in the Network Data Exchange (NDEx), an NIH-approved domain-specific repository for biomolecular networks[35].

RO-Crate data and metadata packages created by FAIRSCAPE clients are first uploaded from each data acquisition laboratory and from the data integration pipeline to the FAIRSCAPE server, where they are assigned ARK identifiers[20]. They are then exported from the server to an instance of Dataverse, which wraps each package in additional standard DataCite schema metadata. More detailed FAIRSCAPE metadata is available in the RO-Crate metadata JSON files associated with each dataset, which are available following RO-Crate download.

## Recommendations

### *Minimum Requirements for AI-Ready Metadata*

The Bridge2AI Standards Working Group (WG) outlined the types of information required to characterize a biomedical dataset as "AI-Ready".[1] This information should primarily be defined in metadata associated with the dataset. It is divided into seven major criteria:

1. **FAIRness**: Digital objects comply with the FAIR Principles.[6]
2. **Provenance:** Origins and transformational history of digital objects are richly documented.
3. **Characterization**: Content semantics, statistics, and standardization of digital objects are well-described. including any quality or bias issues.
4. **Pre-Model Explainability**: Supports explainability of predictions and classifications based on the data with regard to metadata, fit for purpose, and data integrity.[36]
5. **Ethics**: Ethical data acquisition, management, and dissemination are documented and maintained.[37–39]
6. **Sustainability**: Digital objects and their metadata stored in FAIR, stable archives.
7. **Computability**: Standardized, computationally accessible, portable, and contextualized.

Please refer to the AI-Readiness Recommendations article cited above for a detailed set of practices required to adequately support each of the above criteria. We note that currently each Bridge2AI GC has taken an individual and pragmatic approach to implementing these criteria. As of this writing, all GCs are in the process of fully implementing them.

## *Capturing Clinical Metadata*

Clinical data collection in the Bridge2AI GCs builds upon the experiences of numerous researchers, practitioners, and standards developers. The primary value of clinical metadata lies in its ability to provide essential context, transforming raw clinical observations and measurements into meaningful, interpretable information. It captures critical details about patients, diagnoses, treatments, procedures, and outcomes. For instance, metadata can specify the exact protocol used for a blood pressure measurement, the version of a diagnostic code, or the precise criteria for a disease diagnosis, reducing ambiguity and ensuring accurate interpretation. This contextual richness is what makes clinical data truly useful.

To achieve this, clinical metadata must be linked to widely adopted terminologies and data models. Standardized systems such as SNOMED-CT for clinical terms, LOINC for laboratory tests, RxNorm for medications, and ICD-10 for diagnoses ensure consistent meaning across different systems, offering enriched semantics. Data models like the OMOP Common Data Model or HL7 FHIR provide a structured framework for organizing this information, promoting harmonization and interoperability.

**Harnessing Standardized Metadata in Practice**

Large-scale initiatives have demonstrated that aligning technical, ethical, and regulatory frameworks is key to harnessing the power of metadata-rich EHRs. Programs such as the All of Us Research Program[40] and the National Clinical Cohort Collaborative (N3C)[41] have successfully addressed the challenge of resolving differences among data types and sources by adopting standardized frameworks like the OMOP Common Data Model.

These programs also highlight the dual utility of metadata in enhancing care quality while enforcing privacy. By implementing tiered access controls and techniques like geographic generalization and date shifting, they protect sensitive information. Audit trails, for example, enable retrospective analysis of clinical decision patterns and can detect unauthorized access through anomaly monitoring. Similarly, version-controlled documentation preserves a verifiable revision history of clinicians' decision-making processes, which is crucial for compliance and quality audits.

**Applications in Research, AI, and Governance**

*Advanced Analytics and AI:* Clinical metadata is essential for preparing data for advanced Analytics and artificial intelligence (AI). It supports the annotation and labeling of data, providing the necessary context for training AI models for tasks like disease prediction and personalized treatment recommendations. By describing data formats, types, and dictionaries in a standardized way, metadata facilitates the information sharing required for large-scale analysis, such as population health studies and multi-center clinical trials. This interoperability is a major contributor to robust AI performance. Furthermore, metadata helps identify potential biases or limitations in the data, ensuring that healthcare AI algorithms are reliable and equitable.

*Research and Data Discovery*: In research, metadata enhances the discovery of relevant data. It allows researchers to efficiently search for specific patient populations, treatment protocols, or outcomes, and enables flexible, reproducible analysis. Metadata-driven summaries can inform decisions about study feasibility and model design without requiring direct access to raw patient records—a vital feature for privacy-preserving research. The structure defined by metadata also allows for complex queries, describing relationships between different tables in clinical databases to answer sophisticated research questions.

*Data Governance and Compliance*: Metadata is fundamental to data governance and compliance with regulations such as HIPAA and GDPR. It tracks data provenance (origin) and history, creating essential audit trails. Administrative metadata, which covers data ownership, access, permissions, and usage rights, is critical for implementing governance policies and ensuring that the data lifecycle, including retention and archiving, meets regulatory requirements.

**Challenges in Metadata Management**

Despite its importance, improper metadata management presents significant challenges and risks.

*Data Quality and Integrity:* Inaccuracies in data entry, inconsistent formats, and missing information can undermine the performance of AI algorithms. For instance, erroneous timestamps resulting from system glitches or unrecognized time zone shifts can complicate the reconstruction of treatment timelines, affecting time-sensitive clinical decisions. Data captured in different units (e.g., blood pressure in mmHg and kPa) or duplicated entries can fragment the longitudinal view of a patient's health, a problem exacerbated by poor interoperability between the incompatible EHR platforms used by different healthcare providers.

*Privacy and Legal Risks:* While detailed metadata is vital for audits, it can expose sensitive information if not properly secured, necessitating strict data governance. Furthermore, poorly contextualized metadata can have serious downstream consequences. In legal proceedings, detailed information such as timestamps and edit histories could be misinterpreted as evidence of negligence if documentation gaps or normal variations in clinical practice are not properly explained.

Ultimately, EHRs have evolved from simple digital repositories into dynamic tools for clinical decision-making and research. Realizing their full potential to accelerate healthcare innovation demands meticulous attention to metadata. Only when data are accurate, standardized, and secure can they be considered truly "AI-ready."

### *Comprehensive Dataset Release Metadata*

Several attempts have been made in the literature to promote and standardize comprehensive metadata descriptions as single packages. These include Datasheets[42], as well as Data Cards[43], Dataset Nutrition Labels[44], Healthsheets[45], and Croissant[46]. Based on the rapid pace of advances in the AI field and lessons learned from the program, we have determined that the most effective model is a synthesis of these sets of metadata definitions, with the Datasheets model serving as the primary inspiration for a more focused set of computable specifications. The Bridge2AI Standards Working Group initially translated the Datasheets model to the LinkML modeling language[47] as a direct adaptation of the questions posed in Gebru et al.'s 2021 publication, and more recently, with encouragement of GC representatives in the Standards WG, on custom schemas derived from Datasheets but more tightly focused on specific data characteristics and modalities of each GC. Other recent work includes attempts to automatically derive GC-focused LinkML Datasheet schemas from metadata and publications already released by a particular GC.

Concurrently, Bridge2AI GCs evaluated other approaches to standardize project-level metadata. The Salutogenesis GC constructed and released a comprehensive data Healthsheet documenting their data releases. This proved to be a laborious task, although lessons learned from the first effort could likely improve the rapidity and efficiency of constructing other instances. The Functional Genomics GC is working jointly with the Precision Public Health GC to derive Croissant metadata descriptions directly by automation from the existing project metadata. Croissant is directly useful in ML-OPS packages such as ML-FLOW[48,49] and TensorFlow[50], making it a true AI-readiness precursor that may be adopted by modelers with efficiency.

# Next Steps

Standardization of project and dataset-level metadata is well-advanced in the GCs and supported by activities of the Standards WG, but it remains an area of continual improvement. Metadata contents, formats, and infrastructure are actively being harmonized across all GCs. Even with continual progress toward reaching a cohesive data collection across the entirety of Bridge2AI, we recognize that challenges remain for the broader biomedical research community. No single set of practices will serve all groups or purposes equivalently well. This does not invalidate but rather highlights the importance of finding commonalities across approaches to managing metadata. Below, we note efforts within Bridge2AI focused on fostering more unified approaches.

### *Standard Data Description Templates*

Datasheets and Healthsheets were selected as standard data description templates at the project's outset, influenced by NIH requirements. However, since selecting these formats, AI technology and standardization have advanced considerably and have highlighted the need for templates adapted to specific domains and use cases. We have also determined that fully computable templates offer greater reproducibility than written guidelines alone. There are now descriptive templates (including Croissant) that can be directly integrated into machine learning operations pipelines and that may be directly generated by translation from existing metadata. We are actively evaluating how to align these templates with the current Datasheets / Health Sheets model.

### *Cross-GC Metadata Standardization*

Standardizing metadata releases across the Bridge2AI GCs would provide compelling examples of describing experimental results at numerous levels of granularity. By establishing common formats and vocabularies, we can leverage Large Language Models (LLMs) to generate rich metadata annotations, including data provenance, and potentially derive automated explanations for AI models (XAI). This, combined with RO-Crate packaging standards, would enable robust, machine-readable data packages with embedded metadata that can be easily ingested by AI/ML systems. Automated generation of summarized data description templates, directly from existing metadata, would further streamline workflows and enhance AI-readiness for all released data. Having standardized metadata from our four distinct and highly multimodal data generators will provide a uniquely robust resource for developing and validating cross-domain computational techniques.

*Common Fund Data Ecosystem Integration*

The Common Fund Data Ecosystem (CFDE) is a consortium supported by the NIH Common Fund with the goal of enabling broad use of Common Fund data to accelerate discovery. This includes Bridge2AI data. Within the CFDE, the Data Resource Center (DRC) offers opportunities for presenting a unified, searchable collection of Bridge2AI GC metadata extending to the level of individual data entries. Doing so, however, will require alignment of underlying data models with GC needs. Coverage of clinical terms and concepts will be essential. So will mechanisms for ensuring patient privacy: we recommend that a set of minimal, tractable, and applicable selection criteria, which do not impose re-identification risk, be agreed upon between the GCs and the DRC. We also recommend focusing on patient demographics plus a few additional, GC-relevant terms that can be implemented without risk of disclosing or enabling PHI disclosure. Appropriately directed researchers to a GC's data by the DRC, once accredited and licensed to access the data, may then download whatever data the license allows, and perform further cohort selection within that dataset on their own. Having signed off on the data licensing agreements, and subject to its terms, clinical metadata will then be fully under the licensee's control and their site's required PHI security measures. Therefore, patient re-identification risk through metadata exposure on the DRC site would be eliminated.

## Conclusions

This report highlights the progress made in Bridge2AI metadata standardization efforts across the GCs and the Bridge Center, while also identifying areas that require further attention. Addressing these challenges will be crucial to achieving harmonization across datasets and maximizing the benefits of standardization. A common representational model for dataset metadata standardized across all GCs and supporting all recommended AI-Readiness criteria is envisioned and would create significant synergies. Members from across the consortium may pursue these goals through a focus on refining and implementing standardized data description templates, cross-GC metadata standardization, and CFDE DRC integration, ultimately enhancing the efficiency and effectiveness of AI-powered biomedical research.